\documentclass{jpsj2}
\newcommand{\be}{\begin{equation}}
\newcommand{\ee}{\end{equation}}

\newcommand{\bea}{\begin{eqnarray}}
\newcommand{\eea}{\end{eqnarray}}
\newcommand{\bd}{\begin{displaymath}}
\newcommand{\ed}{\end{displaymath}}
\newcommand{\ba}{\begin{array}}
\newcommand{\ea}{\end{array}}
\newcommand{\bi}{\begin{itemize}}
\newcommand{\ei}{\end{itemize}}
\newcommand{\bc}{\begin{center}}
\newcommand{\ec}{\end{center}}
\newcommand{\bfl}{\begin{flushleft}}
\newcommand{\efl}{\end{flushleft}}
\newcommand{\bfr}{\begin{flushright}}
\newcommand{\efr}{\end{flushright}}

\def\6{\partial}

\def\no{\nonumber \\}

\def\={\!\!\!&=&\!\!\!}
\def\+{\!\!\!&&\!\!\!+~}
\def\-{\!\!\!&&\!\!\!-~}

\title{Alternating  Heisenberg  Spin-1/2  Chains in a Transverse Magnetic
Field}

\author{ \textsc{Saeed Mahdavifar$^{1,2}$}\thanks{E-mail address:
mahdavifar@iasbs.ac.ir} and  \textsc{Alireza Akbari$^{2}$
}\thanks{E-mail address: alireza@iasbs.ac.ir}}

\inst{ $^{1}$Department of Physics, Zanjan University,
P.O.Box 313, Zanjan, Iran\\
$^{2}$Institute for Advanced Studies in Basic Sciences, P.O.Box
45195-1159, Zanjan, Iran}

\abst{The ground state phase diagram of the alternating spin-1/2
chains with anisotropic ferromagnetic coupling under the influence
of a symmetry breaking transverse magnetic field is studied. We
have used the exact diagonalization technique. In the limit where
the antiferromagnetic coupling is dominant, we have identified two
Ising-type quantum phase transitions. We have calculated two
critical fields $h_{c_{1}}$  and
 $h_{c_{2}}$, corresponding to the transition between different magnetic phases
of the system. It is found that the intermediate state
($h_{c_{1}}<h<h_{c_{2}}$) is gapful,
 describing  the stripe-antiferromagnetic phase.
}

\kword{Alternating Heisenberg  Chains, Transverse Magnetic Field}

\begin{document}
\maketitle

\section{Introduction}
The effect induced by external magnetic fields in the
   low-dimensional magnets has attracted much interest
   recently from experimental and theoretical points of view.
 In particular, critical properties of the alternating
 spin-1/2 chains in a magnetic field have been a
 field of intense studies.
This seems pertinent in the face of great progress made within the
last years in fabrication of such AF-F
  compounds.
Since, the AF-F chains have a gap in the spin excitation spectrum,
they reveal extremely rich quantum behavior in the presence of
the magnetic field.

A typical example of the ferromagnetic-dominant  AF-F bond
alternating chain is $\rm{(CH_3)_2CHNH_3CuCl_3}$
(isopropylammonium coppertrichloride:
$\rm{IPACuCl_3}$)\cite{manaka97a,manaka98}. The energy gap in the
absence of the external magnetic field is estimated from the
susceptibility to be $17-18\rm{K}$\cite{manaka97a}. This value is
also confirmed by the analysis of the specific
heat\cite{manaka98}. From the viewpoint of the crystal structure,
the origin of the spin gap was expected to be the spin-1/ 2 AF-F
alternating chain along the c-axis\cite{manaka97a}. However, quite
recently, it was suggested that this system should be
characterized as a spin ladder along the a-axis with the strongly
coupled ferromagnetic rungs, namely the antiferromagnetic chain
with effective S = 1, and the excitation gap $\Delta$ was
estimated as $13.6 \rm{K}$ by means of the neutron inelastic
scattering experiments.\cite{Masudao6}

 More recently, it was reported new
inelastic neutron scattering results for $\rm{(CH_3)_2NH_2CuCl_3}$
(Dimethylammonium copper II chloride, also known as:
$\rm{DMACuCl_3}$ or MCCL). The linked-cluster calculations and the
bulk measurements show that $\rm{DMACuCl_3}$ is a realization of
the spin-1/2 alternating AF-F chain by nearly the same strength of
the antiferromagnetic and ferromagnetic couplings.\cite{Stone07}

 There are other examples for the
AF-F alternating spin-1/2 chain
compounds such as: $\rm{[Cu(TIM)]CuCl_4}$
(TIM=2,3,9,10-tetramethyl-1,3,8,10-tetraenecyclo-1,4,8,11-tetraazatetradecane)
and $\rm{(4-BzpipdH)CuCl_3}$
(4-BzpipdH=4-benzylpiperidinium),\cite{Hagiwara97,Takahashi97} and
$CuNb_2O_6$\cite{Kodama99}.

Theoretically, the AF-F alternating chain is expected to have
relation to the Haldane-gap systems\cite{Hal83}, since they are
regarded as the spin-1 antiferromagnetic chain in the large
ferromagnetic coupling limit. The energy gap exist between the
ground state and the first excited state. The spin correlation
function of the ground state decays exponentially as well as
the spin-1 AF chain. The ground state properties\cite{Takada92,
Hida92a, Hida93, Sakai95, Hida92b, Kohmoto92, Yamanaka93} and
low-lying excitations\cite{Hida94} of this model were well
investigated by numerical tools and variational schemes. In
particular, the string order parameter originally defined for the
spin-1 Heisenberg chains\cite{Nijs89} was generalized to this
system. The ground state has the long-range string order, which is
characteristic of the Haldane-gap phase. Hida has shown that the
Haldane phase of the AF-F alternating chain is stable against any
strength of the randomness\cite{Hida99}.

The ground state phase diagram of the AF-F alternating chain in a
magnetic field is studied by the numerical diagonalization and the
finite-size scaling based on the conformal field
theory\cite{Sakai95}. It is shown that the magnetic state is
gapless and described by the Luttinger liquid phase. It is also
found that the magnetic state is characterized by the algebraic
decay of the spin correlation functions. Recently, Yamamoto et. al
described the magnetic properties of the model in a magnetic field
in terms of the spinless fermions and the spin waves\cite{Yamamoto05}.
They employed the Jordan-Wigner transformation and treated the
fermionic Hamiltonian within the Hartree-Fock approximation. They
have also implemented the modified spin wave theory to calculate
the thermodynamic functions as the specific heat and the magnetic
susceptibility.

The effect of an uniform transverse magnetic field on the ground
state phase diagram of a spin-1/2 AF-F chain with anisotropic
ferromagnetic coupling is much less studied. Partly, the reason is
that AF-F alternating chains with anisotropic ferromagnetic
coupling are not still fabricated. However, from the theoretical
point of view these systems are extremely interesting, since they
open a new wide polygon for the study of complicated quantum
behavior, unexpected in the more conventional spin systems. The
Hamiltonian of the model under consideration on a periodic chain
of $N$ sites is given by
\bea
 H & =&  J_{AF}\sum_{j=1}^{N/2}[S_{2j-1}^x S_{2j}^x + S_{2j-1}^yS_{2j}^y
+ S_{2j-1}^zS_{2j}^z]
\no& -& J_F\sum_{j=1}^{N/2} [S_{2j}^xS_{2j+1}^x +
S_{2j}^yS_{2j+1}^y
 +\Delta S_{2j}^zS_{2j+1}^z]
\no& +& h\sum_{j=1}^N
 S_{j}^{x},
 \label{Hamiltonian}
\eea
where $S_{j}^{x, y, z}$ are spin-1/2 operators on the $j$-th site.
$J_{F}$ and $J_{AF}$ denote the ferromagnetic and
antiferromagnetic couplings respectively. The limiting case of
isotropic ferromagnetic coupling corresponds to $\Delta=1$ and $h$
is the transverse magnetic field. The model (1) in the case of
$J_{F}=-J_{AF}$ and $\Delta=1$ corresponds to the isotropic
Heisenberg chain in an external magnetic field. The ground state
phase diagram of this model is known\cite{Dmit02}.

In this paper, we present our numerical results obtained in the
low-energy
 states of the AF-F alternating spin-1/2 chain with anisotropic
ferromagnetic coupling in a transverse magnetic field $h$.
Assuming that the antiferromagnetic coupling is dominant, we study
the effect of a uniform transverse magnetic field on the ground
state phase diagram of the model. In particular, we apply the
modified Lanczos method to diagonalize numerically finite chains.
Using the exact diagonalization results, we calculate the spin
gap, the magnetization, the string order parameter, and various
spin-structure factors as a function of the uniform transverse
magnetic field. Based on the exact diagonalization results we
obtain the magnetic phase diagram of the model showing
Haldane, stripe-antiferromagnetic and ferromagnetic
phases. We denote by "ferromagnetic phase" the phase with the
magnetization parallel to the external field as only the
nonvanishing order parameter. We show that the Haldane phase
remains stable even in the presence of the uniform transverse
field less than a critical field $h_{c_{1}}$.

 The outline of the paper is as follows: In section II we discuss the
model in the strong antiferromagnetic coupling limit and derive
the effective spin chain Hamiltonian to outline the symmetry
aspects of the problem. In section III we present results of the
exact diagonalization calculations using the modified Lanczos
method. Finally we conclude and summarize our results in section
IV.
\begin{figure}[t]
\vspace{1cm} \centering \scalebox{0.35}[0.35]{
\includegraphics*{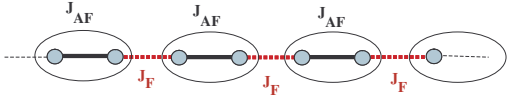}
}
\caption{  Sketch of the Spin-1/2 AF-F alternating Chain
considered in this paper.}
\label{fig1}
\end{figure}


\section{Large Antiferromagnetic Coupling Limit }

In this section we will discuss the model (1) in the limiting case
of the strong antiferromagnetic coupling $J_{AF}\gg J_{F}$. We
will show that in this limit the model can be regarded as a fully
anisotropic XYZ chain, which allows us to outline the symmetry
aspects of the problem under consideration\cite{Japa07}. The
schematic picture of the AF-F alternating chain is plotted in
Fig.~\ref{fig1}. At $J_{AF}\gg J_{F}$, the system behaves as the
nearly independent block of pairs\cite{Mila98}. Indeed an
individual block may be in a singlet
($|S>=1/\sqrt{2}[|\uparrow\downarrow>-|\downarrow\uparrow>]$) or a
triplet state ($|T_{1}>=|\uparrow\uparrow>$,
$|T_{0}>=1/\sqrt{2}[|\uparrow\downarrow>+|\downarrow\uparrow>]$,
$|T_{-1}>=|\downarrow\downarrow>$) with the corresponding energies
given by
$$
E_{1, -1}=\left({J_{AF}\over 4}\ \mp h\right),\,\,
E_0={J_{AF}\over 4},\,\, E_s=-{3J_{AF}\over 4}.
$$
At $h\leq J_{AF}$, one component of the triplet becomes closer to
the singlet ground state such that for a strong enough magnetic
field we have a situation when the singlet and $S^{z} =1$
component of the triplet create a new effective spin $\tau=1/2$
system. One can easily project the original Hamiltonian
(\ref{Hamiltonian}) on the new singlet-triplet subspace
$$
|\Downarrow\rangle  \equiv |s\rangle  =
\frac1{\sqrt2}[|\uparrow\downarrow\rangle -
|\downarrow\uparrow\rangle]\, , \qquad |\Uparrow\rangle  \equiv
|T_{1}\rangle  = |\uparrow\uparrow\rangle \, . \nonumber
$$
When expressed in terms of the effective spin operators $\tau_{j}$
up to the accuracy of irrelevant constant,
 the
original Hamiltonian (\ref{Hamiltonian}) becomes the Hamiltonian
of the spin-1/2 chain fully anisotropic XYZ chain in an effective
magnetic field\cite{Mila98, Totsuka98}
\begin{eqnarray}
H_{\emph{eff}} &=& \frac{J_{F}}{2} \sum_{j=1}^{N/2}
[-\frac{1}{2}\tau^{z}_{j}\tau^{z}_{j+1} + \tau^{y}_{j}
\tau^{y}_{j+1} +\Delta \tau^{x}_{j} \tau^{x}_{j+1}] \nonumber \\
&+& h^{eff} \sum_{j=1}^{N/2} \tau^{z}_{j}\, ,
\label{EffectiveHamiltonian2}
\end{eqnarray}
where $h^{eff} = h - J_{AF} + J_{F}/4$. Note that in deriving
(\ref{EffectiveHamiltonian2}), we have used the rotation in the
effective spin space which interchanges the $x$ and $z$ axes.

At $\Delta = 1$, the effective problem reduces to the theory of
the $XXZ$ chain with a fixed antiferromagneic anisotropy of $1/2$
in a magnetic field. The gapped phase at $h^{eff}  < h^{eff}
_{c_1}$ for the AF-F alternating chain corresponds to the
negatively saturated magnetization phase for the effective spin
chain, whereas the massless phase for the AF-F alternating chain
corresponds to the finite magnetization phase of the effective
spin-1/2 chain. The critical field $h^{eff}_{c_2}$ where the AF-F
alternating chain is totally magnetized, corresponds to the fully
magnetized phase of the effective spin-1/2 chain.

From the exact ground state phase diagram of the anisotropic $XXZ$
chain in a magnetic field \cite{Takahashi99}, it is easy to check
that $h^{eff}_{c_{1},c_{2}}= \mp J_{F}/4$.
 Therefore we can see that the
isotropic AF-F alternating chain in a magnetic field shows a
transition from the Haldane phase to the Luttinger-Liquid phase at
$h_{c_{1}}= J_{AF}-J_{F}/2$ and a transition from the
Luttinger-Liquid  phase into the fully polarized phase at
$h^{ext}_{c_{2}}=J_{AF}$.

Away from the isotropic point $\Delta =1$ the effective
Hamiltonian (\ref{EffectiveHamiltonian2}) describes the fully
anisotropic ferromagnetic XYZ chain in a magnetic field that is
directed perpendicular to the easy axes.  For the particular value
of magnetic field $h^{eff}=0$, the effective XYZ chain is long
range ordered in $"y"-$direction. This is corresponding to the
original AF-F alternating chain being ordered in the direction
perpendicular to the applied magnetic field with opposite
staggered magnetization on the blocks. We have defined this new
phase as a stripe-antiferromagnetic phase (Fig.~\ref{fig5}). The
stripe-antiferromagnetic phase extends over the whole region of
the transverse magnetic field $h_{c_{1}}<h<h_{c_{2}}$. For other
values of the $h^{eff}$ it is clear that the
stripe-antiferromagnetic phase order will be replaced either by
the Haldane phase ($h < h_{c_1}$ correspond to negative effective
fields) or the phase with only one order parameter - magnetization
along the applied field ($h> h_{c_2}$ correspond to positive
effective fields).

\begin{figure}[tbp]
\centerline{\includegraphics[width=8cm,angle=0]{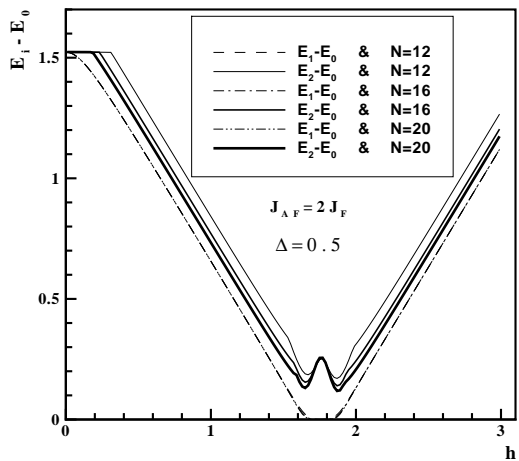}}
\caption{ Difference between the energy of the two lowest levels
and the ground state energy as a function of the transverse
magnetic field, $h$ for $J_{AF}=2J_{F}$ and $\Delta=0.5$ including
different chain lengths $N=12,16,20$. The lowest three lines for
$E_1 - E_0$ are indistinguishable on this scale. }
\label{fig2}
\end{figure}


\section{Numerical Results}

In this section, for exploring the nature of the spectrum and the
phase transition, we use the modified Lanczos method to
diagonalize numerically finite chains ($N=12,16, 20, 24$). The
energies of the few lowest eigenstates were obtained for the
chains with periodic boundary conditions. To find the effect of a
transverse magnetic field on the ground state phase diagram of the
system, we start our consideration by the anisotropic case,
$\Delta\neq1$. First, we have computed the three lowest energy
eigenvalues of $N=12, 16, 20$ chain with $J_{AF}=2 J_{F}$ and
different values of the anisotropy parameter $\Delta=0.25, 0.5,
0.75$. As an example, in Fig.~\ref{fig2} we have plotted results
of this calculations for $\Delta=0.5$. It can be seen that the
difference between the two lowest states, decreases by increasing
the transverse magnetic field. These are independent of the chain
length. Considering  the limiting case of $J_F=0$ the system
consists of the $N/2$ independent block of pairs. Indeed the
energy of the ground state for the $N/2$ blocks is equal to
$N/2\times E_s$, and in the presence of a magnetic field the first
excited state has the energy equal to: $(N-1)/2\times E_s+E_1$.
Thus the deference between two lowest energy of the system will be
independent of the size of the system. However for  the case of
the $J_{AF}\gg J_F$ the other terms in the Hamiltonian behave as a
small perturbation, which causes that
 the three  lines for $E_1 -E_0$ to be indistinguishable on this scale.

In the region of the magnetic fields $h_{c_{1}}< h<h_{c_{2}}$ the
two lowest states form a twofold degenerate ground state in the
thermodynamic limit. Thus, the excitation gap in the system is the
difference between the second excited state and the ground state.
In the case of the zero magnetic field ($h=0$) the spectrum of the
model is gapped. The spectrum remains gapfull except at the two
critical fields:
 $h_{c_{1}}=1.64\pm0.01$ and
$h_{c_{2}}=1.88\pm0.01$. The spin gap, which appears at $h>
h_{c_{1}}$, first increases vs external field, but then starts to
decrease, and finally vanishes at $h_{c_{2}}$. At $h>h_{c_{2}}$
the gap opens again and, for a sufficiently large field becomes
proportional to $h$. These results are in good agreement with the
results obtained in the studies of the fully anisotropic
antiferromagnetic $XYZ$ chain in a magnetic field
\cite{Hogemans05}.

On the other hand, we have also checked
 our numerical tools in the isotropic case, $\Delta=1$. In this case, we
found that the gap decreases linearly and vanishes in the critical
magnetic fields $h_{c_{1}}=1.57\pm0.01$ and $h_{c_{2}}=2.0$, for
the constant coupling $J_{AF}=2 J_{F}$. Both values of the
critical fields are obtained from studying of the finite chains
are very close to the previous section results. Also, our
numerical results showed that the spectrum remains gapless for
$h_{c_{1}}< h<h_{c_{2}}$,  in good agreement with numerical
results reported by Sakai\cite{Sakai95}.

\begin{figure}
\centerline{\includegraphics[width=8cm,angle=0]{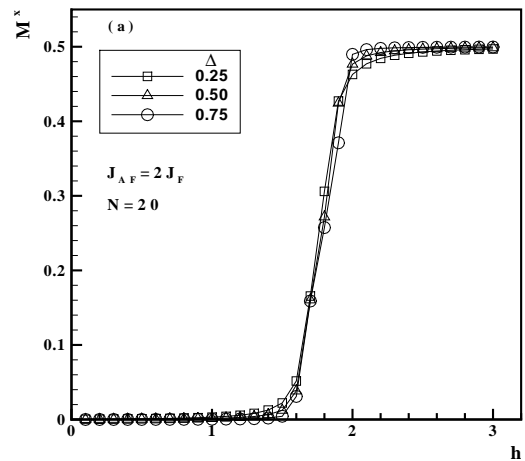}}
\caption{ The transverse magnetization $M_{x}$ as a function of
applied field $h$ for $N=20$ chain with $J_{AF}=2 J_{F}$ and for
different values of the anisotropy parameter $\Delta=0.25, 0.50,
0.75$.
 } \label{fig3}
\end{figure}

To recognize the different phases induced by the transverse
magnetic field in the ground state phase diagram, we have
implemented the modified Lanczos algorithm of finite size chains
$(N=12, 16, 20, 24)$ to calculate the order parameters and the various
spin correlation functions.

The first insight into the nature of the different phases can be
obtained by studying the magnetization along the field axis
\begin{eqnarray}
M^{x}=\frac{1}{N}\sum_{j}\left\langle  S_{j}^{x}\right\rangle,
\end{eqnarray}
where the notation $\langle...\rangle$ represent the expectation
value at the lowest energy state. In Fig.~\ref{fig3} we have
plotted the magnetization along the applied transverse magnetic
field, $M^{x}$, vs $h$ for the chain of length $N=20$. For
arriving at this plot we considered $J_{AF}=2J_{F}$  for the
different values of the anisotropy parameter
$\Delta=0.25,0.5,0.75$. Due to the profound effect of the quantum
fluctuations the transverse magnetization remains small but finite
for $0<h<h_{c_{1}}$ and reaches zero at $h=0$. This is in
agreement with the results of magnetization obtained in the case
of fully anisotropic $XYZ$ chain \cite{Hogemans05}. Increasing the
magnetic field, the magnetization increases
 for $h> h_{c_{1}}$
 very fast. This behavior is in
agreement with expectations, based on general statement that in
the gapped phase, the magnetization along the applied field
appears only at a finite critical value of the magnetic field
equal to the spin gap. However, in finite systems we do not
observe a sharp transition close to the saturation value, which
happens at $h> h_{c_{2}}$. The values of the critical fields
$h_{c_{1}}$ and $h_{c_{2}}$ depend on the anisotropy parameter
$\Delta$. By increasing $\Delta$, the critical fields
 take larger values. Also, our numerical results show  that the
magnetization along the directions perpendicular to the applied
field ($M^y$ and $M^z$) remains zero.

\begin{figure}[tbp]
\centerline{\includegraphics[width=8cm,angle=0]{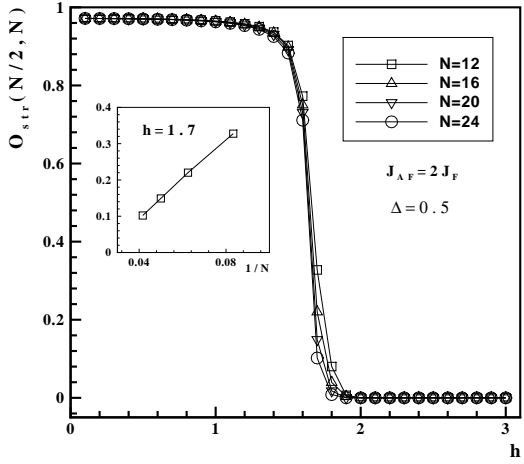}}
\caption{ The string correlation function $O_{str}(l, N)$ for the
$l=N/2$ as a function of the transverse magnetic field, $h$ for
$J_{AF}=2J_{F}$ and $\Delta=0.5$ including different chain lengths
$N=12,16,20, 24$. The inset has shown the
 string order parameter, $O_{str}(N/2, N)$, as a function of
the $1/N$ for a value of magnetic field $h=1.7> h_{c_1}$.}
 \label{fig4}
\end{figure}


We employ the phenomenological renormalization group
(PRG) method\cite{Barber}  to determine these critical
fields ($h_{c_1}$, and  $h_{c_2}$).
In the disordered phase ($h_{c_1} > h$ or $h_{c_2}  < h$), the gap ($E_1- E_0$) tend
to a finite value so that $N(E_1 - E_0)$ increases with $N$. In the Ising
phase $(E_1 - E_0)$ decreases exponentially with $N$ so that $N(E_1 - E_0)$
also decreases with $N$.
The phenomenological renormalization group equation is:

\begin{figure}[tbp]
\centerline{\includegraphics[width=8cm,angle=0]{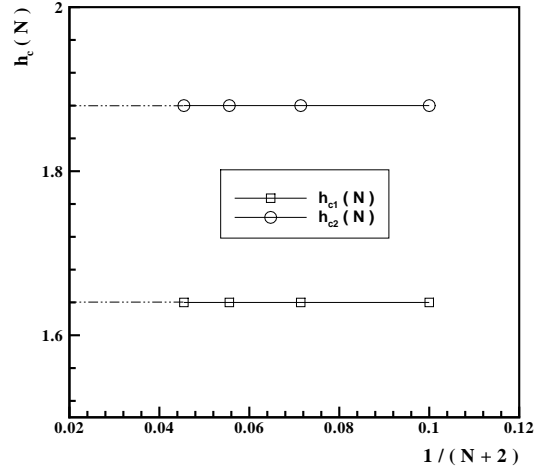}}
\caption{ Fixed points $h_{c_{1}}(N,N +4)$ and $h_{c_{2}}(N,N +4)$ of the phenomenological
 renormalization group equation for the energy gap, plotted versus $1/(N+2)$,
 for $J_{AF}=2J_{F}$ and $\Delta=0.5$ including
different chain lengths $N=8,12,16,20, 24$. The extrapolated values are: $h_{c_1}=1.64\pm
0.01$ and $h_{c_2}=1.88\pm 0.01$. }
\label{prg-fig}
\end{figure}
\begin{eqnarray}
(N+4)g(N+4,h')=Ng(N,h).
\label{prg}
\end{eqnarray}
Where $g(N,h)=E_1(N,h) - E_0(N,h)$ is the gap value for
 chain length $N$ in the magnetic filed $h$.
 At the critical point, $N(E_1 - E_0)$  should be size independent
for large enough systems in which the contribution
from irrelevant operators is negligible. Due to this situation,
we can accurately determine the  critical points by the
PRG method. We can define $h_c(N,N+4)$ as the $N$-dependent fixed point of
Eq.~\ref{prg} and it is extrapolated to the thermodynamic limit,
in order to estimate $h_{c}$.
At the critical point $h=h'=h_{c}$, therefore the curves
of  $N(E_1 - E_0)$ vs. $h$ for sizes $N$ and $N + 4$ cross at certain values
 $h_{c_1}(N,N +4)$ and $h_{c_2}(N,N +
4)$ (’finite size critical points’). The thermodynamic critical points ($h_{c_1}$, and  $h_{c_2}$)
 can be obtained by
appropriately extrapolating $h_{c_1}(N,N +4)$ or $h_{c_2}(N,N + 4)$
to $N\rightarrow \infty$.
Figure~(\ref{prg-fig}) represents the extrapolation procedure of the
transition points, for $J_{AF}=2J_{F}$ and $\Delta=0.5$ including
different chain lengths $N=8,12,16,20, 24$. The values of $h_{c_{1(2)}}(N ,N+4)$
for four pairs of system sizes $(N ,N+4)=
(8,12), (12,16), (16,20)$, and $(20,24)$ are represented by square (circle)
in this figure. The extrapolated values are: $h_{c_1}=1.64\pm 0.01$ and $h_{c_2}=1.88\pm 0.01$.
The system size dependence of the critical points is almost
negligible as shown in Fig.~\ref{prg-fig}.


As we mentioned before, in the absence of the uniform transverse
field, the spectrum is gapful. The ground state of the model is
the Haldane phase with the long-range string order. Thus, the
Haldane phase can be recognized  from studying the string
correlation function. The string correlation function in the chain
of length $N$ defined only for odd $l$ as\cite{Hida99}
\begin{eqnarray}
O_{str}(l, N)=-\left\langle exp\lbrace i \pi
\sum_{k=2j+1}^{2j+1+l} S^{z}_{k}\rbrace \right\rangle.
\end{eqnarray}
In particular we have calculated the string correlation function
$O_{str}(l, N)$ for the $l=N/2$. In Fig.~\ref{fig4} we have
plotted $O_{str}(N/2, N)$ as a function of $h$ for the chain with
$J_{AF}=2J_{F}$, $\Delta=0.5$ and for different values of the
chain length $N=12,16,20,24$. As it can be seen from this figure,
at $h<h_{c_{1}}$, $O_{str}(N/2, N)$ is close to  its maximum value
$1.0$, therefore the chain system is in the Haldane phase. The
Haldane phase remains stable even in the presence of the
transverse field less than $h_{c_{1}}$.

\begin{figure}[t]
\vspace{1cm} \centering \scalebox{0.35}[0.35]{
\includegraphics*{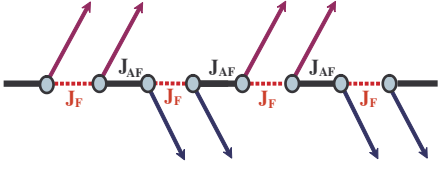}
}
\caption{  Classically, the long-range order canted in the
direction of applied field $h$, which is expected in the ground
state of the spin-1/2 AF-F alternating chain in a transverse
field.}
\label{fig5}
\end{figure}

 In the inset of the Fig.~\ref{fig4} we have also plotted
the string order parameter, $O_{str}(N/2, N)$, as a function of
the $1/N$ for a value of magnetic field $h=1.7> h_{c_{1}}$. It is
clear that by increasing the size of the system, the $O_{str}(N/2,
N)$, converges to the very small values close to zero. Which shows
that there
  is not the string ordering at larger transverse magnetic fields
  $h>h_{c_{1}}$.

\begin{figure}[tbp]
\centerline{\includegraphics[width=8cm,angle=0]{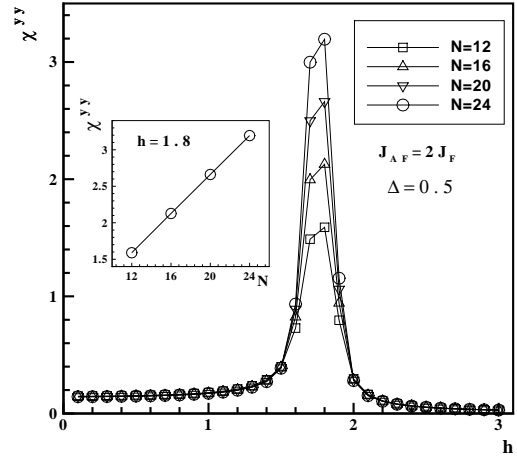}}
\caption{ The correlation function of the stripe-antiferromagnetic
order parameter $\chi^{yy}$ as a function of the transverse
magnetic field, $h$ for $J_{AF}=2J_{F}$ and $\Delta=0.5$ including
different chain lengths $N=12,16,20, 24$. In the inset the
correlation function of the stripe-antiferromagnetic order
$\chi^{yy}$ is plotted versus $N$ for a value of the transverse
field $h_{c_{1}}<h=1.8<h_{c_{2}}$.}
\label{fig6}
\end{figure}

To get additional data about the character of the spin ordering in
the
 intermediate gapped phase, we introduce a new order parameter.
Classically, the effect of the transverse magnetic field $h$ is
interesting.
 Energetically, the ground state of the system has a long-range order canted
 in the direction of the applied transverse magnetic field, as illustrated
in Fig.~\ref{fig5}.
  The ordering of this phase is a kind of the stripe-antiferromagnetic phase.
  Therefore the order parameter
  of the stripe-antiferromagnetic phase has been defined by
\begin{eqnarray}
M_{st}^{y}=\frac{2}{N}\left\langle \sum_{j=1}^{N/2}
(S_{2j-1}^{y}-S^{y}_{2j})(-1)^j\right\rangle.
\end{eqnarray}
For any value of the transverse
 field $h$ in the intermediate phase, the Lanczos results lead
  to $M_{st}^{y}=0$, since the ground state is two-fold  degenerate and in a finite system no
 symmetry breaking happens .
  However the spin correlation function diverges in the ordered phase as
   $N\longrightarrow \infty$. We have computed the correlation function
   of the stripe-antiferromagnetic order parameter given by
\begin{eqnarray}
\chi^{yy}=\left\langle\sum_{n=1}^{N/2}(S_{2j-1}^{y}-S^{y}_{2j})
(S_{2j-1+2n}^{y}-S^{y}_{2j+2n})(-1)^n\right\rangle.
\end{eqnarray}
In Fig.~\ref{fig6} we have plotted $\chi^{yy}$ as a function of
$h$ for the chain with $J_{AF}=2 J_{F}$, $\Delta=0.5$ and for
different values of the chain length $N=12,16,20,24$. As it is
clearly seen from this figure, there is no long rang
stripe-antiferromagnetic order along the $"y"$ direction at
$h<h_{c_{1}}=1.65 \pm 0.05$ and $h> h_{c_{2}} =1.90 \pm 0.05$.
However, in the intermediate region
 $h_{c_{1}}< h <
h_{c_{2}}$, spins of  each block show a profound antiferromagnetic
order in the $"y"$ direction. Thus, between the critical fields
$h_{c_{1}}$ and $h_{c_{2}}$, the ground state of the anisotropic
case is Ising-type symmetry broken phase (stripe-antiferromagnetic
phase) unlike the Luttinger liquid state for the isotropic
case\cite{Sakai95}. In the inset of Fig.~\ref{fig6} we have also
shown the correlation function of the stripe-antiferromagnetic
order $\chi^{yy}$ versus $N$ for a value of transverse field
$h_{c_{1}}<h=1.8<h_{c_{2}}$. It can be seen that the function
$\chi^{yy}$ shows the linear dependence on the chain length $N$,
corresponding to the ordered stripe-antiferromagnetic phase in the
intermediate region $h_{c_{1}}<h<h_{c_{2}}$.

Thus, our numerical results show that the ground state phase
diagram of the antiferromagnetic dominant spin-1/2 AF-F chain in a
transverse field contains, besides the gapped
Haldane and ferromagnetic phases, the
stripe-antiferromagnetic phase. Each phase is characterized by its
own type of long-range order: the ferromagnetic order along the
transverse magnetic field axis in the ferromagnetic phase; the
string order along the $"z"$ axis in the Haldane phase; and the
stripe-antiferromagnetic order along the $"y"$ axis in the
stripe-antiferromagnetic phase.

\section{Conclusions}

In this paper, we have investigated the ground state phase diagram
of the antiferromagnetic dominant ($J_{AF} > J_{F}$) spin-1/2 AF-F
chain with anisotropic ferromagnetic coupling in a transverse
magnetic field $h$. We have implemented the modified Lanczos
method to diagonalize numerically finite chains.
 Using the exact diagonalization results, we have calculated the various order
parameters and the
 spin-structure factors as a function of the transverse magnetic field
$h$. We have found that
  a gapped phase exists in the intermediate values of the transverse field
($h_{c_{1}}<h<h_{c_{2}}$).
  Then, we have identified its ordering as an interesting kind of the
stripe-antiferromagnetic phase. Two quantum phase transitions in
the ground state of the system with increasing transverse magnetic
field have been identified. The first transition corresponds to
the transition from the gapped Haldane phase to the gapped
stripe-antiferromagnetic phase. The other one is the transition
from the gapped stripe-antiferromagnetic phase into the fully
polarized ferromagnetic phase.

On the other hand, in the limit that the antiferromagnetic
coupling is dominant ($J_{AF}\gg J_{F}$) we have mapped the model
(\ref{Hamiltonian}), onto an effective XYZ Heisenberg chain in an
external effective field ($h^{eff}$). This mapping allowed us to
relate the critical fields $h_{c_{1}}$ and $h_{c_{2}}$ to the coupling
constants in the isotropic case $\Delta=1$.

\section*{Acknowledgment}
We would like to thank J. Abouie, R. Jafari and A. Ghorbanzadeh
for insightful comments and  stimulating discussions. We are also
grateful to B. Farnudi and M. Aliee for reading carefully our
manuscript and appreciate their useful comments.


\end{document}